\documentclass[fleqn,english,preprint]{revtex4-1}
\usepackage[T1]{fontenc}
\usepackage[latin9]{inputenc}
\usepackage{amsmath}
\usepackage{amssymb}
\usepackage{stmaryrd}
\usepackage{graphicx}
\usepackage{babel}
\begin{document}
\title{Moment-Fourier approach to ion parallel fluid closures and transport
for a toroidally confined plasma}
\author{Jeong-Young Ji}
\email{j.ji@usu.edu}

\address{Department of Physics, Utah State University, Logan, Utah 84322, USA}
\author{Eric D. Held}
\address{Department of Physics, Utah State University, Logan, Utah 84322, USA}
\author{J. Andrew Spencer}
\address{Department of Physics, Utah State University, Logan, Utah 84322, USA}
\author{Yong-Su Na}
\address{Department of Nuclear Engineering, Seoul National University, Seoul
08826, South Korea}
\begin{abstract}
A general method of solving the drift kinetic equation is developed
for an axisymmetric magnetic field. Expanding a distribution function
in general moments a set of ordinary differential equations are obtained.
Successively expanding the moments and magnetic-field involved quantities
in Fourier series, a set of linear algebraic equations is obtained.
The set of full (Maxwellian and non-Maxwellian) moment equations is
solved to express the density, temperature, and flow velocity perturbations
in terms of radial gradients of equilibrium pressure and temperature.
Closure relations that connect parallel heat flux density and viscosity
to the radial gradients and parallel gradients of temperature and
flow velocity, are also obtained by solving the non-Maxwellian moment
equations. The closure relations combined with the linearized fluid
equations reproduce the same solution obtained directly from the full
moment equations. The method can be generalized to derive closures
and transport for an electron-ion plasma and a multi-ion plasma in
a general magnetic field.
\end{abstract}
\maketitle

\section{Introduction}

For magnetically confined plasmas, neoclassical transport theory describes
particle, heat, and momentum transport of a steady-state plasma due
to Coulomb collisions in an inhomogeneous magnetic field~\citep{Galeev1968S,Rosenbluth1972HH,Hazeltine1973HR,Hinton1976H,Hirshman1981S,Chang1982H,Taguchi1988}.
The neoclassical transport is obtained by solving the first order
drift kinetic equation~\citep{Hazeltine1973,Hazeltine2003M} assuming
a zeroth order background distribution~(see Ref.~\citep{Balescu1988,Helander2002S}
for reviews). Due to difficulty in treating the integro-differential
collision operator in velocity space, modified collision operators
have been adopted for analytical work. Numerical work may adopt the
Landau (Fokker-Planck) collision operator with desired accuracy by
increasing velocity space resolution. Numerous transport codes have
been developed to solve the continuum drift kinetic equation with
a modified~\citep{Belli2008C,Belli2009C} or an exact Landau collision
operator~\citep{Belli2011C,Landreman2012E,Landreman2013E,Held2015e4,Jepson2021e4,Spencer2022e4}.

For describing a macroscopic state of a tokamak plasma, the fluid
variables are of primary importance and solving fluid equations instead
of the kinetic equation may be sufficient. Due to significantly lower
dimensionality of position space compared to phase space, numerically
solving fluid equations has a great advantage over solving the kinetic
equation~\citep{NIMROD2004,M3D2008,M3DC12009,BOUT++2009,JOREK2021}.
The key issue is to obtain proper closures to capture desired physics
effects. Even though the heat flux density is derived in neoclassical
transport theory, it cannot serve as one of closures for the temperature
equation because it is derived from the fluid equations, and hence,
expressed in terms of the zeroth-order density and temperature instead
of the (first-order) fluid variables whose evolution equations are
to be closed. That is, the heat flux derived from the divergence free
condition plays no role for the divergence term in the temperature
equation.

In this work, we introduce an analytic method to solve the drift kinetic
equation to obtain closures and transport. For a magnetized plasma,
the parallel moment equations are derived in Ref.~\citep{Ji2014H}.
One advantage of the moment approach is the availability of the exact
collisional moments of the linearized Landau operator~\citep{Ji2006H}.
The moment-based collision operator can be utilized for the linear
and nonlinear gyrokinetic Coulomb collision operator~\citep{Ji2009H2,Jorge2017RL,Jorge2019FR}.
For slab geometry where the magnetic field strength does not change
along a magnetic field line, the drift-kinetic equation can be converted
to a linear system of ordinary differential equations with constant
coefficients. This linear system can be analytically solved for the
parallel moments using the eigenvector method~\citep{Ji2009HS}. 

On the other hand, for an inhomogeneous magnetic field of a tokamak,
the drift kinetic equations becomes a linear system of ordinary differential
equations with varying coefficients. This means that the eigenvector
method used in the integral closure~\citep{Ji2009HS} does not work.
For a system of linear differential equations with varying coefficients,
we can Fourier-expand the varying coefficients and moments to build
a system of linear algebraic equations. While truncation both in the
moments and Fourier modes is inevitable, the solution of the truncated
system is equivalent to that of the drift kinetic equation when convergence
is achieved by increasing the number of moments and Fourier modes.
The solution moments can then be used to construct the distribution
function that is the solution of the drift kinetic equation. Therefore
the moment solution can be used for benchmarking numerous fluid and
kinetic codes.

In Sec.~\ref{sec:DKEME}, we present the parallel moment equations
which are equivalent to the first order drift kinetic equation. In
Sec. \ref{sec:nuT}, we use the Fourier expansion to solve the general
moment equations for fluid quantities in Fourier series. The convergent
solution is presented as the numbers of moments and Fourier modes
increase. In Sec.~\ref{sec:FE=000026C}, we derive closures and incorporate
them into fluid equations to reproduce the fluid quantities. In Sec.\ref{sec:con},
we conclude and discuss possible extensions of the work to more general
plasmas.

\section{Drift kinetic equation and moment equations\label{sec:DKEME}}

In standard neoclassical transport theory~(see Ref.~\citep{Helander2002S}
for a general review), drift kinetic equations are solved for ion
and electron transport. An analytic solution can be obtained for an
axisymmetric magnetic field 
\begin{equation}
\mathbf{B}=I\nabla\zeta+\nabla\zeta\times\nabla\psi\label{vecB:}
\end{equation}
where $2\pi\psi$ is the poloidal flux, $2\pi I/\mu_{0}$ is the poloidal
current, $\mu_{0}$ is the magnetic permeability, and $\zeta$ is
the toroidal angle. For simplicity, we assume a circular magnetic
field
\begin{equation}
B=\frac{B_{0}}{1+\epsilon\cos\theta}\label{B:}
\end{equation}
where $\theta$ is the poloidal angle, $B_{0}$ is a constant reference
field, $\epsilon=r/R_{0}$ is the inverse aspect ratio, and $R_{0}$
and $r$ respectively are the major and minor radii of a circular-shape
flux surface.

For ion transport, the ion-electron collisions are often ignored and
the reduced ion drift kinetic equation for the first-order distribution
function $f_{1}$ becomes

\begin{equation}
v_{\|}\partial_{\|}(f_{1}-F)=C(f_{1})\label{dke1}
\end{equation}
with 
\begin{equation}
F=-\frac{Iv_{\|}}{\Omega}\frac{df_{0}}{d\psi}=-\frac{Iv_{\|}}{\Omega}\left[\frac{d\ln p_{0}}{d\psi}+\left(s^{2}-\frac{5}{2}\right)\frac{d\ln T_{0}}{d\psi}\right]f_{0}\label{F:}
\end{equation}
and
\begin{equation}
f_{0}(\psi,w)=\frac{n_{0}(\psi)}{\left[2\pi mT_{0}(\psi)\right]^{3/2}}e^{-w/T_{0}(\psi)}=\frac{n_{0}}{\pi^{3/2}v_{0}^{3}}e^{-s^{2}}\label{f0}
\end{equation}
in the $(\psi,\theta,w=mv^{2}/2,\mu=mv_{\perp}^{2}/2B)$ coordinates,
where $\partial_{\|}=\mathbf{b}\cdot\nabla=(\mathbf{B}/B)\cdot\nabla$,
$v_{\|}=\mathbf{b}\cdot\mathbf{v}$, $\Omega=qB/m$, $v_{0}=\sqrt{2T_{0}/m}$,
and $s=v/v_{0}$. Note that flux surfaces can be labeled by the lowest-order
density $n_{0}$, temperature $T_{0}$, or pressure $p_{0}=n_{0}T_{0}$.
The collision operator is a Landau operator linearized with respect
to a static Maxwellian distribution function $f_{0}$, 
\begin{equation}
C(f_{1})=C(f_{1},f_{0})+C(f_{0},f_{1}).\label{C1}
\end{equation}

One difficulty of solving the kinetic equation (\ref{dke1}) is in
treating the collision operator, an integro-differential operator
in velocity space. In standard analytical neoclassical theory, the
Landau operator is often approximated as the Lorentz pitch-angle scattering
operator with an additional momentum restoring term for an analytical
treatment. In the moment approach, the linearized collision operator
can be analytically calculated and explicitly represented by a matrix
of collision coefficients. In this work, we solve a system of parallel
moment equations introduced in Ref.~\citep{Ji2006H,Ji2014H}. The
moment equations can also be derived from the drift kinetic equation
as shown below.

In the moment method of this work, a gyro-averaged distribution function
$f_{1}$ is expanded as 
\begin{equation}
f_{1}=f_{0}\sum_{l,k}\hat{P}^{lk}\hat{M}^{lk}\label{fex}
\end{equation}
with orthonormal polynomials
\[
\hat{P}^{lk}=\frac{1}{\sqrt{\bar{\sigma}_{lk}}}P^{lk}=\frac{1}{\sqrt{\bar{\sigma}_{lk}}}s^{l}P_{l}(v_{\|}/v)L_{k}^{(l+1/2)}(s^{2}),
\]
where $P_{l}$ is a Legendre polynomial, $L_{k}^{(l+1/2)}$ is an
associated Laguerre (Sonine) polynomial, and the normalization constants
are 
\begin{equation}
\bar{\sigma}_{lk}=\bar{\sigma}_{l}\lambda_{lk},\;\bar{\sigma}_{l}=\frac{1}{2l+1},\;\lambda_{lk}=\frac{(l+k+1/2)!}{k!(1/2)!}.\label{norm}
\end{equation}
Several lowest order moments of $f_{1}$ are: $\hat{M}^{00}=n_{1}/n_{0}$
(density), $\hat{M}^{01}=-\sqrt{3/2}T_{1}/T_{0}$ (temperature), $\hat{M}^{10}=\sqrt{2}u/v_{0}$
(parallel flow velocity $u=V_{1\|}$), $\hat{M}^{11}=-\sqrt{4/5}h_{\|}/v_{0}p_{0}$
(parallel heat flux density), and $\hat{M}^{20}=\sqrt{3/4}\pi_{\|}/p_{0}$
(parallel viscosity), where $p_{0}=n_{0}T_{0}$. The neoclassical
thermodynamic drive term can also be expanded as
\begin{align}
v_{\|}\partial_{\|}F= & \frac{v_{0}\partial_{\|}\ln B}{B/B_{0}}f_{0}\Biggl[\left(2\hat{P}^{00}-2\sqrt{\frac{2}{3}}\hat{P}^{01}+\frac{1}{\sqrt{3}}\hat{P}^{20}\right)\hat{p}_{0,\psi}\nonumber \\
 & +\left(-5\sqrt{\frac{2}{3}}\hat{P}^{01}+2\sqrt{\frac{10}{3}}\hat{P}^{02}+\frac{1}{\sqrt{3}}\hat{P}^{20}-\sqrt{\frac{7}{6}}\hat{P}^{21}\right)\hat{T}_{0,\psi}\Biggr],\label{vdF}
\end{align}
where 
\begin{align}
\hat{p}_{0,\psi} & =\frac{I}{qv_{0}B_{0}n_{0}}\frac{dp_{0}}{d\psi},\label{p0psi}\\
\hat{T}_{0,\psi} & =\frac{I}{qv_{0}B_{0}}\frac{dT_{0}}{d\psi}.\label{T0psi}
\end{align}

Taking the $\hat{P}^{jp}$ moment of Eq.~(\ref{dke1}) yields
\begin{equation}
\sum_{lk}\psi^{jp,lk}\partial_{\|}\hat{M}^{lk}+\psi_{B}^{jp,lk}(\partial_{\|}\ln B)\hat{M}^{lk}=\frac{1}{\lambda_{\mathrm{C}}}c^{jp,lk}\hat{M}^{lk}+\frac{\partial_{\|}\ln B}{B/B_{0}}\left(g_{p}^{jp}\hat{p}_{0,\psi}+g_{T}^{jp}\hat{T}_{0,\psi}\right),\label{ME1}
\end{equation}
where $\lambda_{\mathrm{C}}=v_{0}\tau_{\mathrm{ii}}$ (the ion mean
free path). Note that eliminating $(j,p)=(0,0)$, $(0,1)$, and (1,0)
moment equations from Eq.~(\ref{ME1}) yields a set of closure moment
equations, similar to the closure moment equations in slab geometry
Ref.~\citep{Ji2017LH}. The constant coefficients $\psi^{jp,lk}$,
$\psi_{B}^{jp,lk}$, and $c^{jp,lk}$ are defined by 
\begin{equation}
\int\mathrm{d}^{3}vv_{\|}\hat{P}^{jp}\hat{P}^{lk}f_{0}=n_{0}v_{0}\psi^{jp,lk},\label{:ps}
\end{equation}
\begin{equation}
\int\mathrm{d}^{3}vv_{\|}\hat{P}^{jp}(\partial_{\|}\hat{P}^{lk})f_{0}=n_{0}v_{0}(\partial_{\|}\ln B)\psi_{B}^{jp,lk},\label{:psB}
\end{equation}
\begin{equation}
\int\mathrm{d}^{3}v\hat{P}^{jp}C(f_{0}\hat{P}^{lk})=\frac{n_{0}}{\tau_{\mathrm{ii}}}c^{jp,lk}=\frac{n_{0}}{\tau_{\mathrm{ii}}}\delta_{jl}c_{pk}^{j}.\label{:c}
\end{equation}
The nonvanishing $g^{jp}$ in Eq.~(\ref{vdF}) are 
\begin{equation}
g_{p}^{0,0}=2,\;g_{p}^{0,1}=-2\sqrt{\frac{2}{3}},\;g_{p}^{2,0}=\frac{1}{\sqrt{3}}\label{gp:}
\end{equation}
and
\begin{equation}
g_{T}^{0,1}=-5\sqrt{\frac{2}{3}},\;g_{T}^{0,2}=2\sqrt{\frac{10}{3}},\;g_{T}^{2,0}=\frac{1}{\sqrt{3}},\;g_{T}^{2,1}=-\sqrt{\frac{7}{6}}.\label{gT:}
\end{equation}

Noting that $\psi^{jp,lk}=\delta_{j,j\pm1}\psi_{lk}^{j\pm}$, $\psi_{B}^{jp,j+1,k}=-(j+2)\psi^{jp,j+1,k}/2$,
and $\psi_{B}^{jp,j-1,k}=(j-1)\psi^{jp,j-1,k}/2$ (see Ref.~\citep{Ji2014H})
and defining
\begin{align}
\partial_{\|}^{j+} & =\partial_{\|}-\frac{j+2}{2}\partial_{\|}\ln B,\nonumber \\
\partial_{\|}^{j-} & =\partial_{\|}+\frac{j-1}{2}\partial_{\|}\ln B,\label{dpm}
\end{align}
we can combine the $\psi$ and $\psi_{B}$ terms to rewrite Eq.~(\ref{ME1})
as
\begin{equation}
\sum_{k}\psi_{pk}^{j-}\partial_{\|}^{j-}\hat{M}^{j-1,k}+\sum_{k}\psi_{pk}^{j+}\partial_{\|}^{j+}\hat{M}^{j+1,k}=\frac{1}{\lambda_{\mathrm{C}}}\sum_{k}c_{pk}^{j}\hat{M}^{jk}+\frac{\partial_{\|}\ln B}{B/B_{0}}\left(g_{p}^{jp}\hat{p}_{\psi}+g_{T}^{jp}\hat{T}_{\psi}\right).\label{ME1pm}
\end{equation}
Although Eq.~(\ref{ME1}) for $j=0,1,\cdots,L-1$ and $k=0,1,\cdots,K-1$
is a truncated system, there exist $L$ and $K$ such that the solution
does not change when increasing the number of moments higher than
$L$ and $K$. In other words, there exists a convergent solution
of Eq.~(\ref{ME1}) which can be considered as a solution of Eq.~(\ref{dke1}).
Therefore Eq.~(\ref{ME1}) for the truncated set of moments is quantitatively
equivalent to Eq.~(\ref{dke1}).

\section{Fourier method of solving moment equations\label{sec:nuT}}

In the axisymmetric magnetic field (\ref{vecB:}), physical quantities
on a flux surface depends on $\theta$ only. Using $\partial_{\|}=(\mathbf{B}\cdot\nabla\theta/B)\partial/\partial\theta=(B^{\theta}/B)\partial_{\theta}$
and dividing Eq.~(\ref{ME1}) by $B^{\theta}/B$ yields a system
of ordinary differential equations 
\begin{equation}
\sum_{lk}\psi^{jp,lk}\partial_{\theta}\hat{M}^{lk}+\psi_{B}^{jp,lk}(\partial_{\theta}\ln B)\hat{M}^{lk}=\frac{B}{B^{\theta}\lambda_{\mathrm{C}}}c^{jp,lk}\hat{M}^{lk}+\frac{\partial_{\theta}\ln B}{B/B_{0}}\left(g_{p}^{jp}\hat{p}_{\psi}+g_{t}^{jp}\hat{T}_{\psi}\right).\label{MEth}
\end{equation}
Since the coefficient $\partial_{\theta}\ln B$ is $\theta$-dependent,
the eigenvector method used in deriving integral closures~\citep{Ji2009HS}
does not work. Instead, we adopt the Fourier method to convert the
system of differential equations to a system of algebraic equations.
Note that Eq.~(\ref{MEth}) forms a linear system of ordinary differential
equations for the parallel moments $\hat{M}^{lk}$ and the Fourier
expansion of coefficients, moments, and drive terms will convert the
differential system to a linear algebraic system.

In the Fourier method, all physical quantities are expanded in Fourier
series. For $A=\hat{M}^{lk}(\theta)$ and $\partial_{\theta}\ln B/(B/B_{0})$,
\begin{equation}
A(\theta)=A_{(0)}+A_{(1-)}\sin\theta+A_{(1+)}\cos\theta+A_{(2-)}\sin2\theta+A_{(2+)}\cos2\theta+\cdots=\sum_{m}A_{(m)}\varphi_{(m)},\label{a:Fo}
\end{equation}
with Fourier modes 
\begin{multline}
\varphi_{(0)}=1,\;\varphi_{(1)}=\varphi_{(1-)}=\sin\theta,\;\varphi_{(2)}=\varphi_{(1+)}=\cos\theta,\cdots,\\
\varphi_{(2n-1)}=\varphi_{(n-)}=\sin n\theta,\;\varphi_{(2n)}=\varphi_{(n+)}=\cos n\theta,\cdots\label{phi:}
\end{multline}
where the Fourier index is denoted in the parentheses. The Fourier
coefficient for $A(\theta)$ can be obtained by
\begin{equation}
A_{(m)}=\frac{1}{\sigma_{(m)}}\int d\theta\varphi_{(m)}A(\theta),\label{A(m):}
\end{equation}
where $\sigma_{(0)}=2\pi$ and $\sigma_{(m)}=\pi$ for $m>0$. The
derivative $\partial_{\theta}$ and the $\theta$-dependent coefficients
in Eq.~(\ref{MEth}) become matrices in Fourier representation. For
$O=\partial_{\theta},$ $\partial_{\theta}\ln B,$ and $B/B^{\theta}\lambda_{\mathrm{C}}$,
the Fourier matrix elements $O_{(i,j)}$ are obtained by
\begin{equation}
O_{(i,j)}=\frac{1}{\sigma_{(i)}}\int d\theta\varphi_{(i)}O\varphi_{(j)},\label{O(i,j):}
\end{equation}
and the Fourier representation of $O\hat{M}^{lk}$ becomes
\begin{equation}
\left(O\hat{M}^{lk}\right)_{(i)}=\frac{1}{\sigma_{(i)}}\int d\theta\varphi_{(i)}O\sum_{j}\hat{M}_{(j)}^{lk}\varphi_{(j)}=\sum_{j}O_{(i,j)}\hat{M}_{(j)}^{lk}.\label{OM(i):}
\end{equation}
Then the $(m)$th Fourier component of Eq.~(\ref{MEth}) becomes
a system of algebraic equations
\begin{multline}
\psi^{jp,lk}\left(\partial_{\theta}\right)_{(m,n)}\hat{M}_{(n)}^{lk}+\psi_{B}^{jp,lk}\left(\partial_{\theta}\ln B\right)_{(m,n)}\hat{M}_{(n)}^{lk}=\\
c^{jp,lk}\left(\frac{B}{B^{\theta}\lambda_{\mathrm{C}}}\right)_{(m,n)}\hat{M}_{(n)}^{lk}+\left(\frac{\partial_{\theta}\ln B}{B/B_{0}}\right)_{(m)}\left(g_{p}^{jp}\hat{p}_{0,\psi}+g_{T}^{jp}\hat{T}_{0,\psi}\right),\label{MFE}
\end{multline}
where summation over $l,k,$ and $n$ is implied. The system of algebraic
equations can be written in matrix form, 
\begin{equation}
\left\llbracket \psi\partial_{\theta}\right\rrbracket \left\llbracket \hat{M}\right\rrbracket +\left\llbracket \psi_{B}\partial_{\theta}\ln B\right\rrbracket \left\llbracket \hat{M}\right\rrbracket =\left\llbracket cB/B^{\theta}\lambda_{\mathrm{C}}\right\rrbracket \left\llbracket \hat{M}\right\rrbracket +\left\llbracket (g_{p}\hat{p}_{0}+g_{T}\hat{T}_{0})(B_{0}/B)(\partial_{\theta}\ln B)\right\rrbracket ,\label{aM=00003Dg}
\end{equation}
where $\left\llbracket \psi\partial_{\theta}\right\rrbracket =\left[\psi\right]\otimes\left(\partial_{\theta}\right)_{\mathtt{F}}$,
$\left\llbracket \psi_{B}\partial_{\theta}\ln B\right\rrbracket =\left[\psi_{B}\right]\otimes\left(\partial_{\theta}\ln B\right)_{\mathtt{F}}$,
and $\left\llbracket cB/B^{\theta}\lambda_{\mathrm{C}}\right\rrbracket =\left[c\right]\otimes\left(B/B^{\theta}\lambda_{\mathrm{C}}\right)_{\mathtt{F}}$
with $\otimes$ denoting a tensor product of two matrices. The $i$th
row and $j$th column of a Fourier matrix $\left(O\right)_{\mathtt{F}}$
is $O_{(i,j)}$, and the dimension of the linear system is $N=LKF=$
(the number of Legendre polynomials)(the number of Laguerre polynomials)(the
number of Fourier modes).

\begin{figure}
\includegraphics{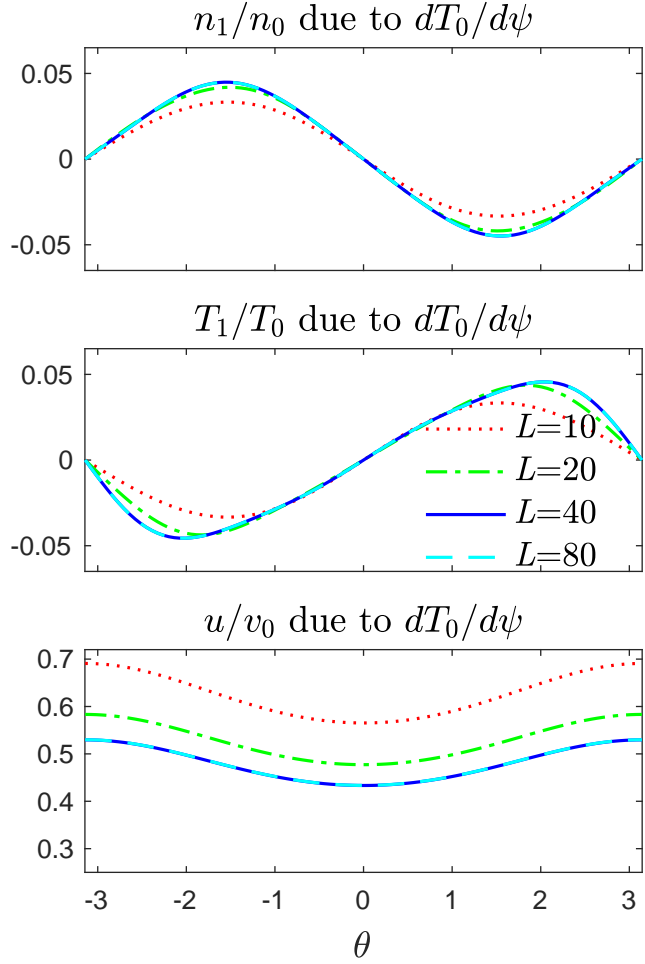}\caption{First-order density, temperature, and parallel flow velocity for $\epsilon=0.1,$
$\mathtt{K}_{0}=100$, $\mathtt{n_{F}}=4$, and for $LK=10\times20$
(red, dotted), $20\times40$ (green, dash-dotted), $40\times80$ (blue
solid), and $80\times160$ (cyan, dashed). The ratios $n_{1}/n_{0}$,
$T_{1}/T_{0}$, and $u/v_{0}$ are plotted in units of $\hat{T}_{0,\psi}$.}
\label{fig:e1p2m}
\end{figure}
\begin{figure}
\includegraphics{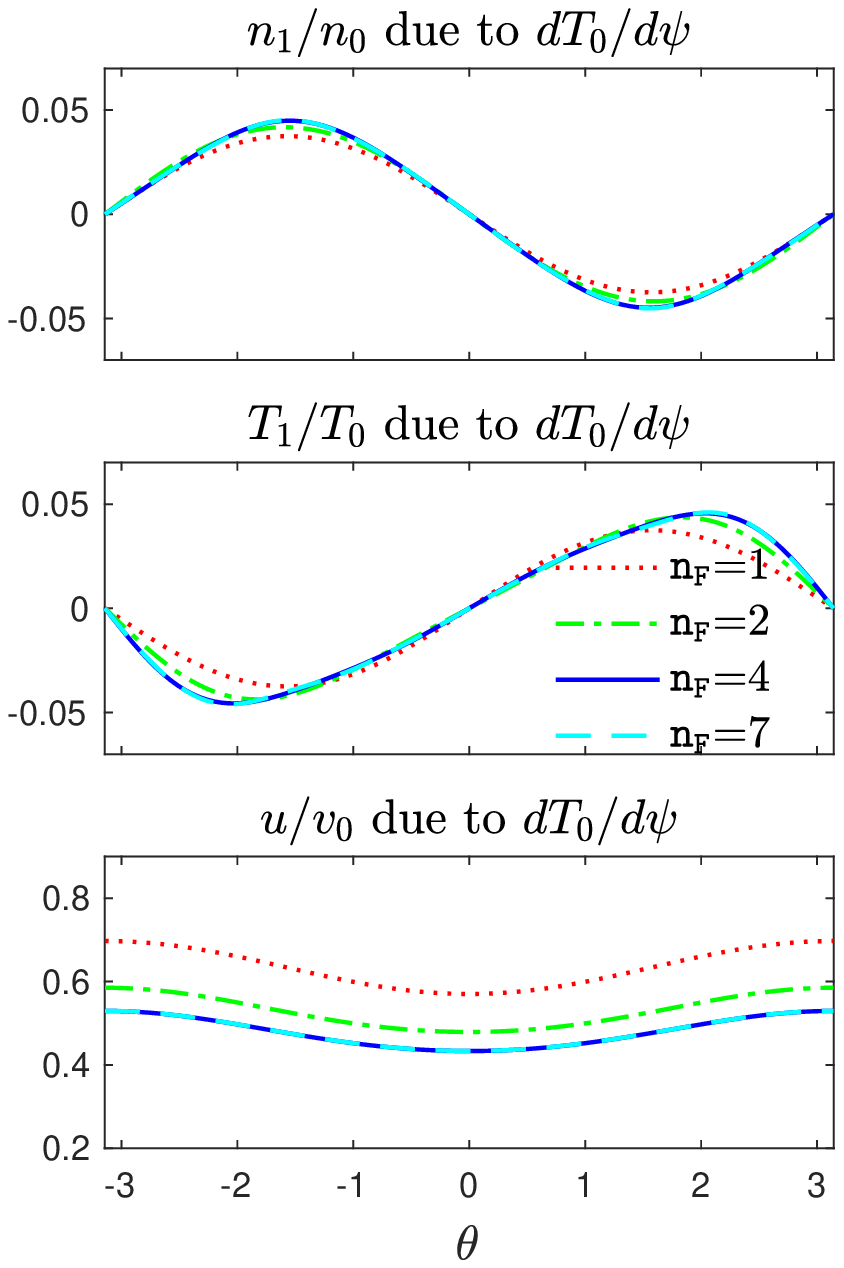}\caption{First-order density, temperature, and parallel flow velocity for $\epsilon=0.1,$
$\mathtt{K}_{0}=100$, $LK=40\times80$, and for $\mathtt{n_{F}}=1$
(red, dotted), 2 (green, dash-dotted), $4$ (blue solid), and $7$
(cyan, dashed). The ratios $n_{1}/n_{0}$, $T_{1}/T_{0}$, and $u/v_{0}$
are plotted in units of $\hat{T}_{0,\psi}$.}
\label{fig:e1p2f}
\end{figure}
\begin{figure}
\includegraphics{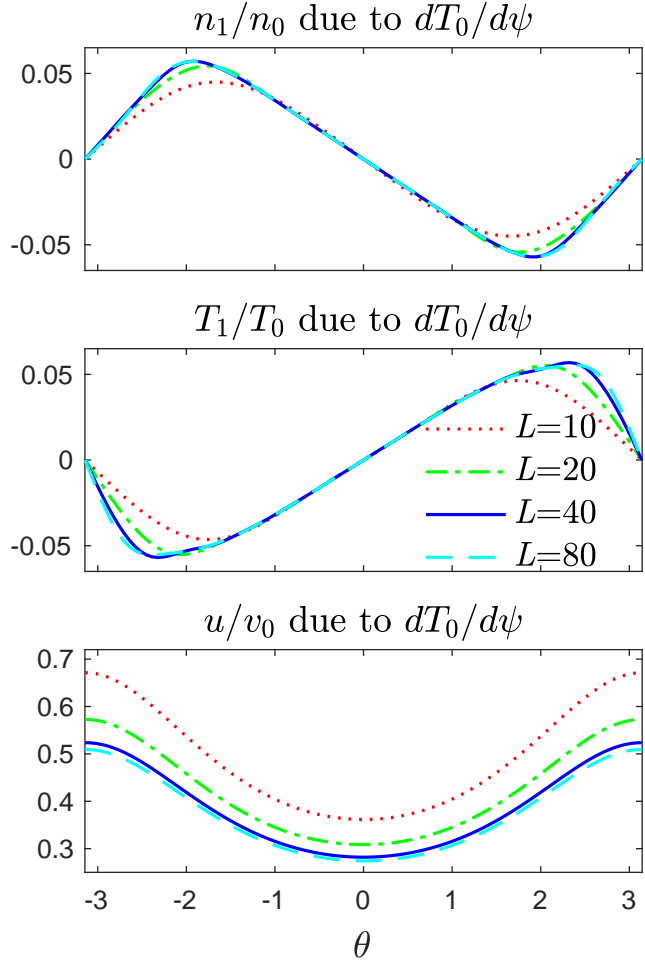}\caption{First-order density, temperature, and parallel flow velocity for $\epsilon=0.3,$
$\mathtt{K}_{0}=100$, $\mathtt{n_{F}}=4$, and for $LK=10\times20$
(red, dotted), $20\times40$ (green, dash-dotted), $40\times80$ (blue
solid), and $80\times160$ (cyan, dashed).}
\label{fig:e3p2m}
\end{figure}
\begin{figure}
\includegraphics{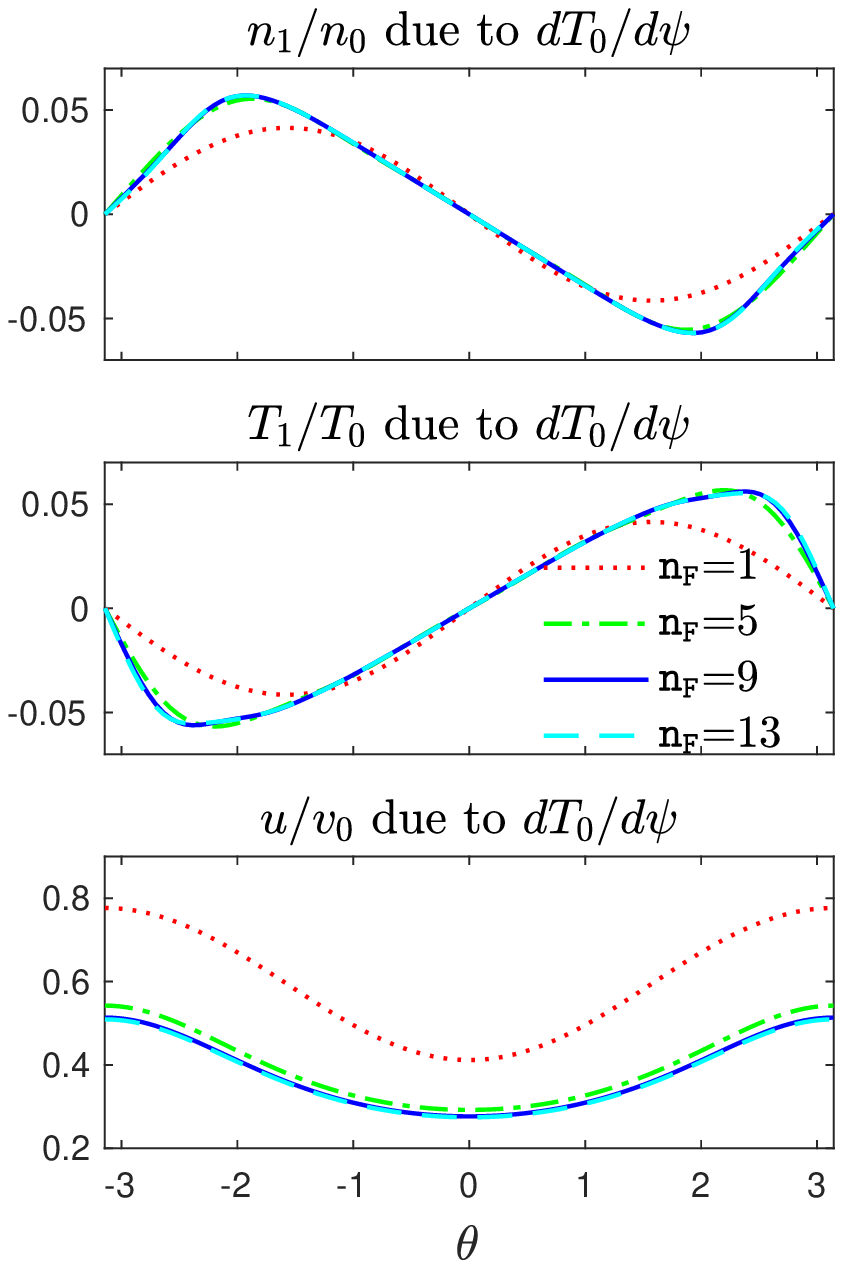}\caption{First-order density, temperature, and parallel flow velocity for $\epsilon=0.3,$
$\mathtt{K}_{0}=100$, $N=LK=40\times80$, and for $\mathtt{n_{F}}=1$
(red, dotted), 5 (green, dash-dotted), $9$ (blue solid), and $13$
(cyan, dashed). The ratios $n_{1}/n_{0}$, $T_{1}/T_{0}$, and $u/v_{0}$
are plotted in units of $\hat{T}_{0,\psi}$.}
\label{fig:e3p2f}
\end{figure}
\begin{figure}
\includegraphics[scale=0.48]{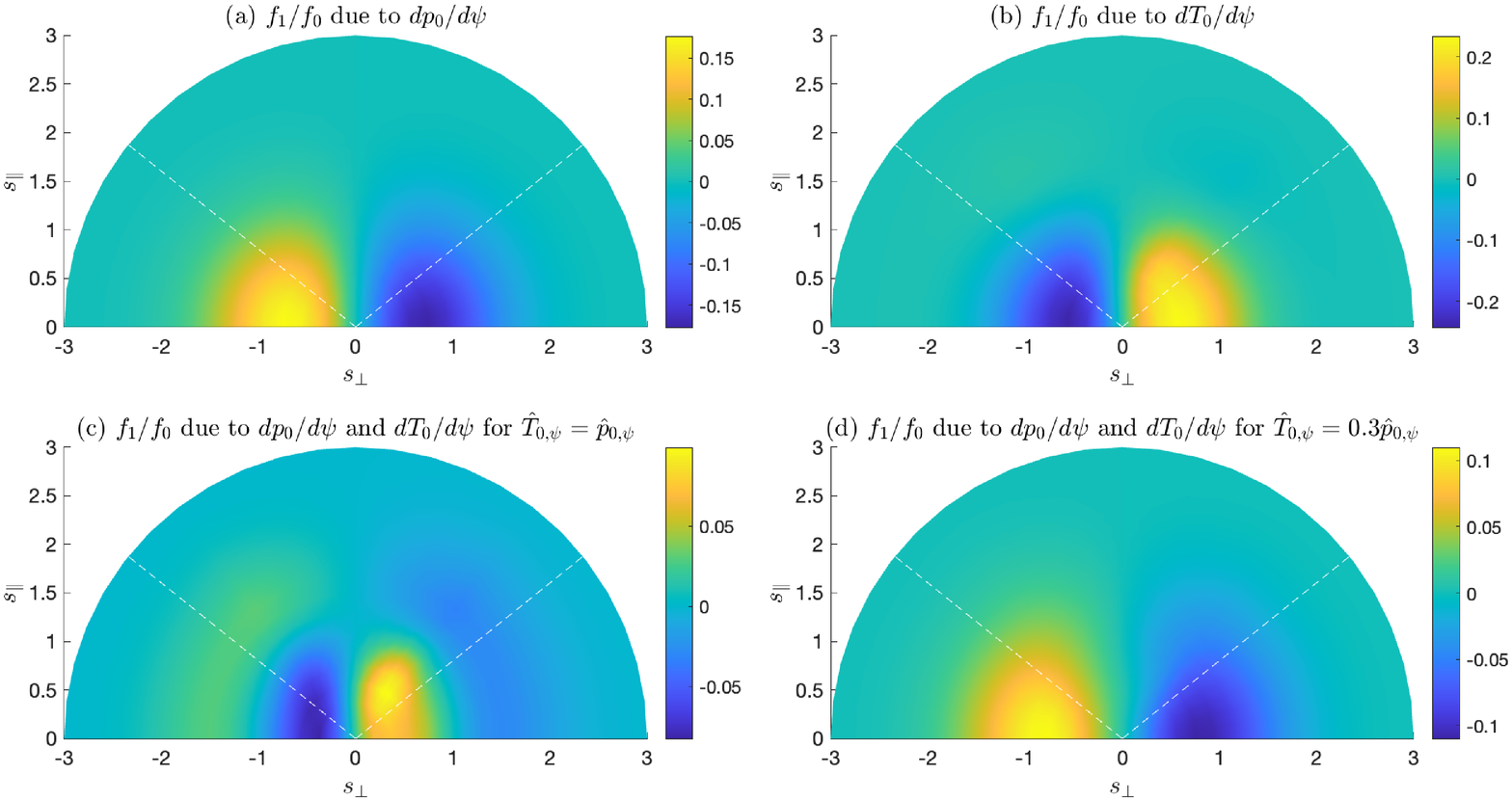}\caption{The first-order distribution function $f_{1}$ at $\theta=-\pi/3$
in the $s_{\perp}$-$s_{\|}$ plane for $\epsilon=0.3$ and $\mathtt{K}_{0}=100$.
The white dashed lines indicate the passing/trapped boundary. The
ratio $f_{1}/f_{0}$ is plotted in units of $\hat{p}_{0,\psi}$ in
(a), (c), and (d) and in units of $\hat{T}_{0,\psi}$ in (b).}
\label{fig:fslsp-}
\end{figure}
\begin{figure}
\includegraphics[scale=0.48]{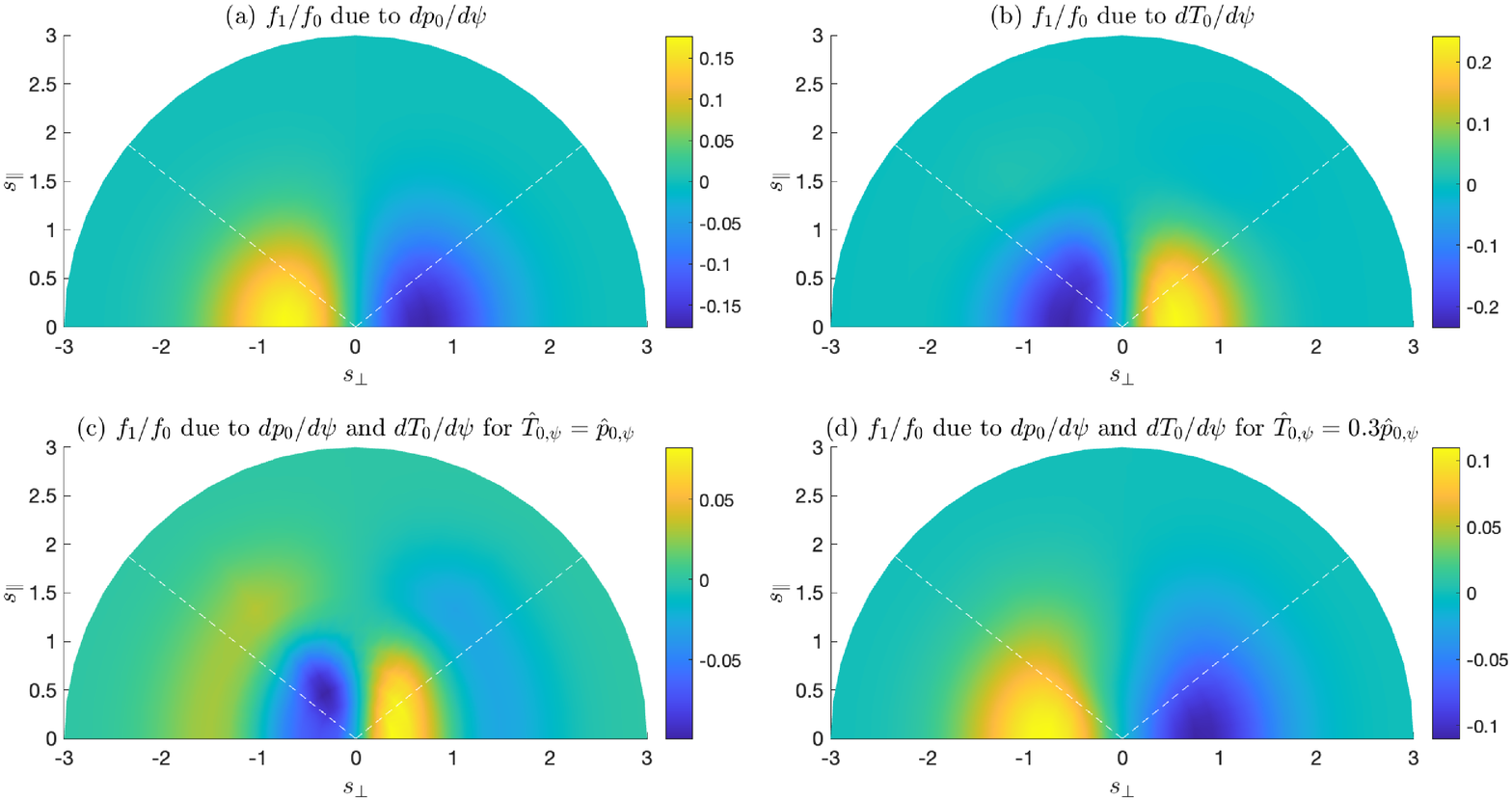}\caption{The first-order distribution function $f_{1}$ at $\theta=\pi/3$
on the $s_{\perp}$-$s_{\|}$ plane for $\epsilon=0.3$ and $\mathtt{K}_{0}=100$.
The white dashed lines indicate the passing/trapped boundary. The
ratio $f_{1}/f_{0}$ is plotted in units of $\hat{p}_{0,\psi}$ in
(a), (c), and (d) and in units of $\hat{T}_{0,\psi}$ in (b).}
\label{fig:fslsp+}
\end{figure}
\begin{figure}
\includegraphics[scale=0.52]{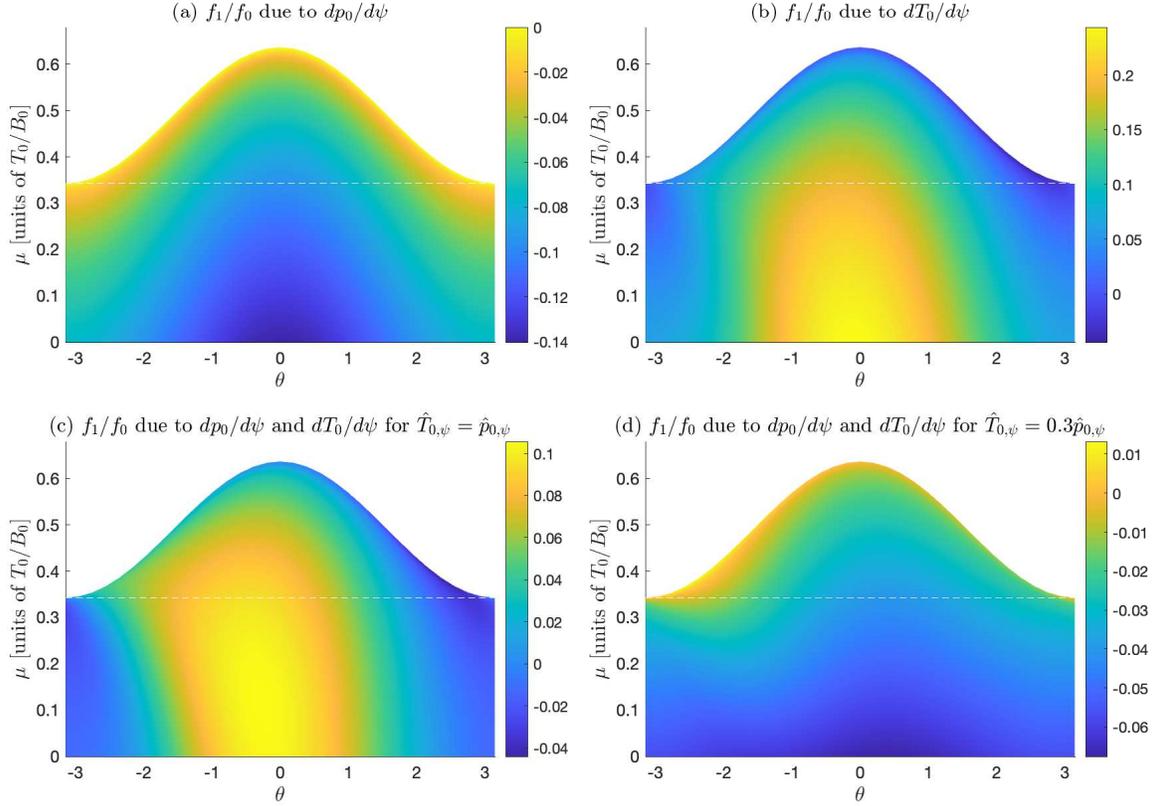}\caption{The first-order distribution function $f_{1}$ at $s=0.7$ on the
$\theta$-$\mu$ plane for $\epsilon=0.3$ and $\mathtt{K}_{0}=100$.
The white dashed line indicates the passing/trapped boundary. The
ratio $f_{1}/f_{0}$ is plotted in units of $\hat{p}_{0,\psi}$ in
(a), (c), and (d) and in units of $\hat{T}_{0,\psi}$ in (b).}
\label{fig:fthmu}
\end{figure}
The solution $\left\llbracket \hat{M}\right\rrbracket $ can be obtained
by inverting or singular-value-decomposing the matrix, 
\begin{equation}
\left\llbracket \hat{M}\right\rrbracket =\left(\left\llbracket \psi\partial_{\theta}\right\rrbracket +\left\llbracket \psi_{B}\partial_{\theta}\ln B\right\rrbracket -\left\llbracket cB/B^{\theta}\lambda_{\mathrm{C}}\right\rrbracket \right)_{\mathrm{ns}}^{-1}\left\llbracket (g_{p}\hat{p}_{0,\psi}+g_{T}\hat{T}_{0,\psi})(B_{0}/B)(\partial_{\theta}\ln B)\right\rrbracket ,\label{M:sol}
\end{equation}
where the subscript `ns' denotes the nonsingular part of the matrix.
It is found that eliminating $n_{(0)}$ and $T_{(0)}$ components
makes the matrix nonsingular {[}see also remarks in relation to Eqs.~(\ref{DT:})
and (\ref{DP:}){]}. Then the Fourier components of the first order
fluid quantities can be read from the solution $\left\llbracket \hat{M}\right\rrbracket $,
\begin{align}
\mathbb{N} & =\hat{p}_{0,\psi}\mathbb{N}^{p_{0}}+\hat{T}_{0,\psi}\mathbb{N}^{T_{0}},\nonumber \\
\mathbb{T} & =\hat{p}_{0,\psi}\mathbb{T}^{p_{0}}+\hat{T}_{0,\psi}\mathbb{T}^{T_{0}},\label{NTU:}\\
\mathbb{U} & =\hat{p}_{0,\psi}\mathbb{U}^{p_{0}}+\hat{T}_{0,\psi}\mathbb{U}^{T_{0}},\nonumber 
\end{align}
where $\mathbb{N}=(\hat{n})_{\mathtt{F}}=(n_{1}/n_{0})_{\mathtt{F}},$
$\mathbb{T}=(\hat{T})_{\mathtt{F}}=(T_{1}/T_{0})_{\mathtt{F}},$ $\mathbb{U}=(\hat{u})_{\mathtt{F}}=(u/v_{0})_{\mathtt{F}}$,
$\mathbb{N}^{\alpha}$, $\mathbb{T}^{\beta}$, and $\mathbb{U}^{\beta}$
$(\beta=p_{0},T_{0})$ are column vectors of Fourier components. With
the Fourier components, the first-order fluid quantities can be constructed
from Eq.~(\ref{a:Fo}). For example, the density due to $\hat{p}_{0,\psi}$
and $\hat{T}_{0,\psi}$, respectively, are $\hat{n}=\sum_{m}\mathbb{N}_{(m)}^{p_{0}}\varphi_{(m)}\hat{p}_{0,\psi}$
and $\hat{n}=\sum_{m}\mathbb{N}_{(m)}^{T_{0}}\varphi_{(m)}\hat{T}_{0,\psi}$,
where $\mathbb{N}_{(m)}^{\beta}$ is the $(m)$th Fourier component
of the column vector $\mathbb{N}^{\beta}$. 

The inverse collisionality of the system is characterized by a Knudsen
number, the ratio of the mean free path to the gradient scale length.
Defining a basic Knudsen number for a tokamak $\mathtt{K}_{0}=B/B^{\theta}\lambda_{\mathrm{C}}$,
the effective Knudsen number would be roughly $\mathtt{K}_{0}\partial_{\theta}\ln B\sim m\mathtt{K}_{0}$
where $m$ is the typical Fourier mode of the system. Although the
solution (\ref{M:sol}) can be obtained for an arbitrary axisymmetric
magnetic field, circular magnetic fields {[}see Eq.~(\ref{B:}){]}
are considered in this work. For the circular magnetic field (\ref{B:}),
the basic Knudsen number is given by $\mathtt{K}_{0}\sim\lambda_{\mathrm{C}}/qR_{0}$
where $q$ is the safety factor and the Fourier mode $m$ is determined
by the inverse aspect ratio $\epsilon=r/R_{0}$. In general, the effective
Knudsen number increases as $\lambda_{\mathrm{C}}$ and $\epsilon$
increase. 

The solution responding to the radial pressure gradient $dp_{0}/d\psi$
shows that $\mathbb{N}^{p_{0}}=0$, $\mathbb{T}^{p_{0}}=0$, and $\mathbb{U}^{p_{0}}=-(1,0,\epsilon,\cdots)^{\mathtt{T}}=-(B_{0}/B)_{\mathtt{F}}$.
This means that the $\hat{p}_{0,\psi}$ drive contributes only to
the flow velocity as $\hat{u}=-\hat{p}_{0,\psi}B_{0}/B+\gamma^{u}B/B_{0}$,
consistent with the continuity equation $\nabla\cdot(n_{0}\mathbf{V}_{1})=0$.
Here $\gamma^{u}$ is an integration constant that can be determined
by temperature and flow velocity equations. It turns out that $\gamma^{u}$
is proportional to $\hat{T}_{0,\psi}$ as verified from the solution
and as discussed in Sec.~\ref{sec:FE=000026C}.

For the solution responding to the radial temperature gradient $dT_{0}/d\psi$,
the density, temperature, and parallel flow velocity are shown in
Fig.~\ref{fig:e1p2m} in the case of $\epsilon=0.1$, $\mathtt{K_{0}}=100$,
and $\mathtt{n_{F}}=4$ $(F=2\mathtt{n_{F}}+1=9)$. A convergence
study increases the number of moments to show that the $LK=40\times80$
moment solution converges and can be considered practically exact.
Note that the polynomials $\hat{P}^{lk}$ in Eq.~(\ref{fex}) form
a complete set. The necessary number of moments for convergence increases
as $\mathtt{K_{0}}$ increases. A convergence study that increases
the number of Fourier modes from 1 to 7 (see Figure \ref{fig:e1p2f})
shows that the $\mathtt{n_{F}}=4$ mode solution converges and may
be considered to be very accurate. The necessary number of Fourier
modes for convergence increases as $\epsilon$ increases. 

Figures \ref{fig:e3p2m} and \ref{fig:e3p2f} show the density, temperature,
and parallel flow velocity for $\epsilon=0.3$, a larger inverse aspect
ratio, and $\mathtt{K_{0}}=100$. The $LK=40\times80$ moment solution,
while not as accurate as in the $\epsilon=0.1$ case, is still very
accurate for practical use, and the $LK=80\times160$ solution is
expected to be accurate. This is because $\epsilon=0.3$ requires
more Fourier modes than $\epsilon=0.1$ for an accurate expansion
of the magnetic field. Higher Fourier modes make the effective Knudsen
number larger. The necessary number of Fourier modes for convergence
is $\mathtt{n_{F}}=13$. 

The moment solution can be used to construct the distribution function
that is a solution of the kinetic equation (\ref{dke1}). Since all
fluid quantities relevant to physical observables involve several
lowest order of moments, the reconstruction of the distribution function
from the moments may be redundant. Nevertheless, the distribution
function itself is important for understanding the kinetic behavior
of a plasma. In the moment expansion, the high order moments near
truncation of the moment expansion could be inaccurate and may adversely
affect the convergence of the distribution function. However we find
that those moments near truncation are several orders smaller than
the fluid moments, making the truncation errors ignorable once the
convergence is achieved. Figures \ref{fig:fslsp-} and \ref{fig:fslsp+}
show the distribution functions constructed from the moment solution
on the $s_{\perp}$-$s_{\|}$ plane at $\theta=-\pi/3$ and $\pi/3$,
respectively. Figure \ref{fig:fthmu} shows the distribution function
at $s=0.7$ on the $\theta$-$\mu$ plane.

\section{Fluid equations and closures\label{sec:FE=000026C}}

In neoclassical transport theory, one solves Eq.~(\ref{dke1}) to
express $f_{1}$ in terms of $f_{0}$ (or $F$) and take moments of
the solution $f_{1}$ to express $u$ in terms of $dp_{0}/d\psi$
and $dT_{0}/d\psi$. These expressions can be directly obtained by
solving Eq.~(\ref{ME1}). In this section we derive closure relations
that can be used for closing and advancing (nonlinear) fluid equations
for density, flow velocity, and temperature. They can also be incorporated
into linearized fluid equations to reproduce the expressions of $n_{1},T_{1}$
and $u$ that are obtained in Sec.~\ref{sec:nuT}. Although the closures
are represented in the Fourier basis, the formalism developed here
can be applied to any basis such as a finite element basis or finite
difference basis in numerical methods.

The linearized fluid equations for $n_{1},$ $u,$ and $T_{1}$ can
be obtained from the original fluid equations with $n=n_{0}+n_{1}$,
$T=T_{0}+T_{1}$, $\mathbf{V}=u\mathbf{b}+\mathbf{b}\times\nabla p_{0}/n_{0}qB$,
$\mathbf{h}=h_{\|}\mathbf{b}+5p_{0}\mathbf{b}\times\nabla T_{0}/2qB$,
and $\boldsymbol{\pi}=(3\pi_{\|}/2)(\mathbf{b}\mathbf{b}-b^{2}\mathsf{I}/3)$
where $\mathbf{b}=\mathbf{B}/B$. They are equivalent to the $\{P^{00},mv_{0}P^{10},-T_{0}P^{01}\}$
moments of Eq.~(\ref{dke1}) and can be read from Eq.~(\ref{MEth})
for $(j,p)=(0,0),$ $(1,0),$ and $(0,1)$:
\begin{equation}
\partial_{\theta}^{0+}\hat{u}=2\hat{p}_{0,\psi}\frac{\partial_{\theta}\ln B}{B/B_{0}},\label{me00}
\end{equation}
\begin{equation}
\partial_{\theta}^{0+}\hat{u}+\partial_{\theta}^{0+}\hat{h}=(2\hat{p}_{0,\psi}+5\hat{T}_{0,\psi})\frac{\partial_{\theta}\ln B}{B/B_{0}},\label{me01}
\end{equation}
\begin{equation}
\partial_{\theta}^{1-}\hat{n}+\partial_{\theta}^{1-}\hat{T}+\partial_{\theta}^{1+}\hat{\pi}=0,\label{me10}
\end{equation}
where $\hat{u}=u/v_{0}$, $\hat{h}=h_{\|}/v_{0}p_{0}$, $\hat{\pi}=\pi_{\|}/p_{0}$,
and $\partial_{\theta}^{l\pm}$ is defined by Eq.~(\ref{dpm}) with
$\partial_{\|}$ replaced by $\partial_{\theta}$. For this fluid
system to be closed, closure quantities $\hat{h}$ and $\hat{\pi}$
should relate to first-order ($\hat{n},$ $\hat{u},$ and $\hat{T}$)
and equilibrium ($\hat{p}_{0,\psi}$ and $\hat{T}_{0,\psi}$) fluid
quantities. 

\begin{figure}
\includegraphics[scale=0.8]{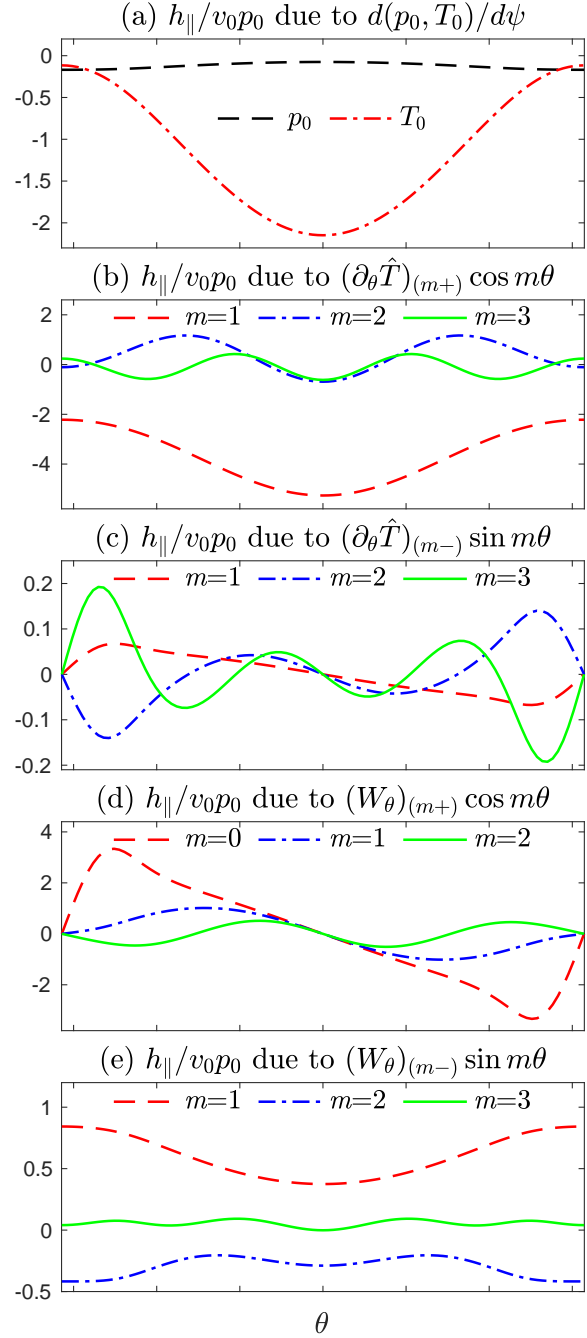}\caption{Parallel heat flux density due to (a) $dp_{0}/d\psi$ and $dT_{0}/d\psi$,
(b) $(\partial_{\theta}\hat{T})_{(m+)}\cos m\theta$, (c) $(\partial_{\theta}\hat{T})_{(m-)}\sin m\theta$,
(d) $(W_{\theta})_{(m+)}\cos m\theta$, and (e) $(W_{\theta})_{(m-)}\sin m\theta$.
The dimensionless heat flux, $h_{\|}/v_{0}p_{0}$, is plotted in units
of (a) $\hat{p}_{0,\psi}$ and $\hat{T}_{0,\psi}$, (b) $(\partial_{\theta}\hat{T})_{(m+)}$,
(c) $(\partial_{\theta}\hat{T})_{(m-)}$, (d) $(W_{\theta})_{(m+)}$,
and (e) $(W_{\theta})_{(m-)}$.}
\label{fig:cih}
\end{figure}

\begin{figure}
\includegraphics[scale=0.8]{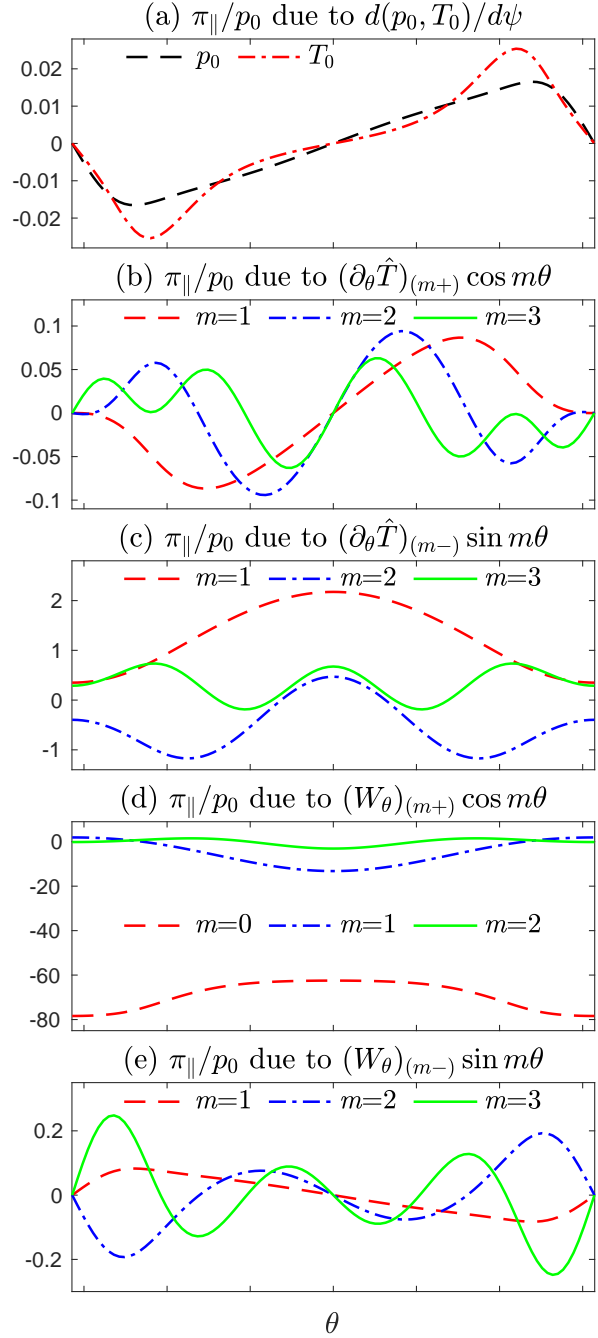}\caption{Parallel viscosity due to (a) $dp_{0}/d\psi$ and $dT_{0}/d\psi$,
(b) $(\partial_{\theta}\hat{T})_{(m+)}\cos m\theta$, (c) $(\partial_{\theta}\hat{T})_{(m-)}\sin m\theta$,
(d) $(W_{\theta})_{(m+)}\cos m\theta$, and (e) $(W_{\theta})_{(m-)}\sin m\theta$.
The dimensionless viscosity $\pi_{\|}/p_{0}$ is plotted in units
of (a) $\hat{p}_{0,\psi}$ and $\hat{T}_{0,\psi}$, (b) $(\partial_{\theta}\hat{T})_{(m+)}$,
(c) $(\partial_{\theta}\hat{T})_{(m-)}$, (d) $(W_{\theta})_{(m+)}$,
and (e) $(W_{\theta})_{(m-)}$.}
\label{fig:cip}
\end{figure}

In order to obtain the closure relations, the rows corresponding to
fluid equations need to be removed from Eq.~(\ref{MEth}). Then the
corresponding columns appear as drives (sources) $\left[g_{\theta}\right]$
in the system: 
\begin{equation}
\left[\psi^{\prime}\right]\left[\partial_{\theta}\hat{M}^{\prime}\right]+\left[\psi_{B}^{\prime}\right](\partial_{\theta}\ln B)\left[\hat{M}^{\prime}\right]=\frac{B}{B^{\theta}\lambda_{\mathrm{C}}}\left[c^{\prime}\right]\left[\hat{M}^{\prime}\right]+\left[g_{\theta}\right]+\frac{\partial_{\theta}\ln B}{B/B_{0}}\left(\left[g_{p}^{\prime}\right]\hat{p}_{0,\psi}+\left[g_{T}^{\prime}\right]\hat{T}_{0,\psi}\right),\label{ME1'}
\end{equation}
where $\prime$ denotes the removal of fluid columns and rows. For
example, $\left[\hat{M}^{\prime}\right]$ is a column vector $(\hat{M}^{0,2},\cdots\hat{M}^{0,K+1},\hat{M}^{1,1},\cdots,\hat{M}^{1,K},\hat{M}^{2,0},\cdots,\hat{M}^{2,K-1},\cdots,\hat{M}^{L-1,0},\cdots,\hat{M}^{L-1,K-1})$.
The nonvanishing elements of $\left[g_{\theta}\right]$ are
\begin{align}
g_{\theta}^{1,1} & =\frac{\sqrt{5}}{2}\partial_{\theta}\hat{T},\label{gth11}\\
g_{\theta}^{2,0} & =-\frac{\sqrt{3}}{2}W_{\theta},\;W_{\theta}=\frac{4}{3}\partial_{\|}^{2-}\hat{u}.\label{gth20}
\end{align}
From Fourier representation of Eq.~(\ref{ME1'}),
\begin{equation}
\left\llbracket \psi^{\prime}\partial_{\theta}\right\rrbracket \left\llbracket \hat{M}^{\prime}\right\rrbracket +\left\llbracket \psi_{B}^{\prime}\partial_{\theta}\ln B\right\rrbracket \left\llbracket \hat{M}^{\prime}\right\rrbracket =\left\llbracket cB/B^{\theta}\lambda_{\mathrm{C}}\right\rrbracket \left\llbracket \hat{M}^{\prime}\right\rrbracket +\left\llbracket g_{\theta}\right\rrbracket +\left\llbracket (g_{p}^{\prime}\hat{p}_{0}+g_{T}^{\prime}\hat{T}_{0})(B_{0}/B)(\partial_{\theta}\ln B)\right\rrbracket ,\label{aM'=00003Dg}
\end{equation}
the solution can be obtained,
\begin{equation}
\left\llbracket \hat{M}^{\prime}\right\rrbracket =\left(\left\llbracket \psi^{\prime}\partial_{\theta}\right\rrbracket +\left\llbracket \psi_{B}^{\prime}\partial_{\theta}\ln B\right\rrbracket -\left\llbracket cB/B^{\theta}\lambda_{\mathrm{C}}\right\rrbracket \right)^{-1}\left\llbracket g_{\theta}+(g_{p}\hat{p}_{0,\psi}+g_{T}\hat{T}_{0,\psi})(B_{0}/B)(\partial_{\|}\ln B)\right\rrbracket .\label{M':sol}
\end{equation}
Fourier components of closures $\hat{h}=-\sqrt{5}\hat{M}^{1,1}/2$
and $\hat{\pi}=2\hat{M}^{2,0}/\sqrt{3}$ can be read from the solution
and expressed in terms of $\hat{p}_{0,\psi}$, and $\hat{T}_{0,\psi}$,
$\hat{T}$, and $\hat{u}$: 
\begin{align}
\mathbb{H} & =\hat{p}_{0,\psi}\mathbb{H}^{p_{0}}+\hat{T}_{0,\psi}\mathbb{H}^{T_{0}}+\mathsf{K}^{hh}\mathsf{D}\mathbb{T}+\mathsf{K}^{h\pi}\mathbb{W},\label{HF:}\\
\mathbb{S} & =\hat{p}_{0,\psi}\mathbb{S}^{p_{0}}+\hat{T}_{0,\psi}\mathbb{S}^{T_{0}}+\mathsf{K}^{\pi h}\mathsf{D}\mathbb{T}+\mathsf{K}^{\pi\pi}\mathbb{W},\label{SF:}
\end{align}
where $\mathbb{H}=(\hat{h})_{\mathtt{F}},$ $\mathbb{S}=\left(\hat{\pi}\right)_{\mathtt{F}}$,
and $\mathbb{W}=\left(W_{\theta}\right)_{\mathtt{F}}=(4/3)\mathsf{D}^{2-}\mathbb{U}\equiv\mathsf{D}_{W}\mathbb{U}$,
$\mathbb{H}^{\beta}$, and $\mathbb{S}^{\beta}$ $(\beta=p_{0},T_{0})$
are column vectors, and $\mathsf{D}=\left(\partial_{\theta}\right)_{\mathtt{F}},$
$\mathsf{D}^{l\pm}=(\partial_{\theta}^{l\pm})_{\mathtt{F}},$ and
$\mathsf{K}^{\alpha\beta}$ $(\alpha,\beta=h,\pi)$ are matrices.
Here a column vector $\mathbb{H}^{\beta}$ and $\mathbb{S}^{\beta}$
connects the closures $h_{\|}$ and $\pi_{\|}$ to a radial gradient
of zeroth-order pressure $(\beta=p_{0})$ or temperature $(\beta=T_{0})$,
and a matrix $\mathsf{K}^{\alpha\beta}$ connects closures $\alpha=h$
and $\pi$ to a parallel gradient of first-order temperature $(\beta=h)$
or parallel flow velocity $(\beta=\pi)$. The closures in the position
space can be constructed from the solution vector, for example, $\hat{h}(\theta)=\sum_{i}\varphi_{(i)}\{\mathbb{H}_{(i)}^{p_{0}}\hat{p}_{0,\psi}+\mathbb{H}_{(i)}^{T_{0}}\hat{T}_{0,\psi}+\sum_{j}[\mathsf{K}_{(i,j)}^{hh}(\mathsf{D}\mathbb{T})_{(j)}+\mathsf{K}_{(i,j)}^{h\pi}\mathbb{W}_{(j)}]\varphi_{(j)}\}$,
where $\mathbb{H}_{(i)}^{\beta}$ is the $(i)$th Fourier component
of the column vector $\mathbb{H}^{\beta}$ and $\mathsf{K}_{(i,j)}^{\alpha\beta}$
is the $(i)$th row and $(j)$th column of the matrix $\mathsf{K}^{\alpha\beta}$.
Figures \ref{fig:cih} and \ref{fig:cip}, respectively, show the
parallel heat flux density and viscosity due to $\hat{p}_{0,\psi}$,
$\hat{T}_{0,\psi}$, and several Fourier modes of $\partial_{\theta}\hat{T}$
and $W_{\theta}$. As the Fourier mode of the thermodynamic drives
increases, the contribution to the closure quantity decreases.

By combining closure relations with the time-independent, linear fluid
equations, we can reproduce the fluid variables of Sec.~\ref{sec:nuT}.
Using $(B_{0}/B)\partial_{\theta}\ln B=-\partial_{\theta}(B_{0}/B)$
and eliminating Eq.~(\ref{me00}) from Eq.~(\ref{me01}), we write
the Fourier representation of Eqs.~(\ref{me00})-(\ref{me10}),
\begin{equation}
\mathsf{D}^{0+}\mathbb{U}=-2\hat{p}_{0,\psi}\mathsf{D}\mathbb{B}_{-1},\label{00F}
\end{equation}
\begin{equation}
\mathsf{D}^{0+}\mathbb{H}=-5\hat{T}_{0,\psi}\mathsf{D}\mathbb{B}_{-1},\label{01F}
\end{equation}
\begin{equation}
\mathsf{D}\mathbb{N}+\mathsf{D}\mathbb{T}+\mathsf{D}^{1+}\mathbb{S}=0,\label{10F}
\end{equation}
where $\mathbb{B}_{-1}=\left(B_{0}/B\right)_{\mathtt{F}}$. Then we
combine with closures (\ref{HF:}) and (\ref{SF:}) to write
\begin{equation}
\mathsf{L}\left(\begin{array}{c}
\mathbb{N}\\
\mathbb{T}\\
\mathbb{U}
\end{array}\right)=\mathbb{R}^{p_{0}}\hat{p}_{0,\psi}+\mathbb{R}^{T_{0}}\hat{T}_{0,\psi}.\label{LNTU:}
\end{equation}
where
\begin{equation}
\mathsf{L}=\left(\begin{array}{ccc}
0 & 0 & \mathsf{D}^{0+}\\
0 & \mathsf{D}^{0+}\mathsf{K}^{hh}\mathsf{D} & \mathsf{D}^{0+}\mathsf{K}^{h\pi}\mathsf{D}_{W}\\
\mathsf{D} & \mathsf{D}+\mathsf{D}^{1+}\mathsf{K}^{\pi h}\mathsf{D} & \mathsf{D}^{1+}\mathsf{K}^{\pi\pi}\mathsf{D}_{W}
\end{array}\right),\label{L:}
\end{equation}
\begin{equation}
\mathbb{R}^{p_{0}}=-\left(\begin{array}{c}
2\mathsf{D}\mathbb{B}^{-1}\\
\mathsf{D}^{0+}\mathbb{H}^{p_{0}}\\
\mathsf{D}^{1+}\mathbb{S}^{p_{0}}
\end{array}\right),\;\mathbb{R}^{T_{0}}=-\left(\begin{array}{c}
0\\
5\mathsf{D}\mathbb{B}^{-1}\\
\mathsf{D}^{1+}\mathbb{S}^{T_{0}}
\end{array}\right).\label{XA:}
\end{equation}
Using the singular value decomposition, we can invert the nonsingular
part of $\mathsf{L}$ and obtain the solution vector $(\mathbb{N},\mathbb{T},\mathbb{U})$
in terms of $\hat{p}_{0,\psi}$ and $\hat{T}_{0,\psi}$. The solution
vector reproduces Eq.~(\ref{NTU:}) with the column vector $(\mathbb{N}^{\beta},\mathbb{T}^{\beta},\mathbb{U}^{\beta})=\left(\mathsf{L}_{\mathrm{ns}}^{-1}\right)\mathbb{R}^{\beta}$
for $\beta=p_{0}$ and $T_{0}$.

Now we discuss how to obtain the parallel flow velocity and heat flux
density when not using the singular value decomposition but instead,
analytically calculating the integration constants. From Eqs.~ (\ref{00F})
and (\ref{01F}), we have
\begin{align}
\mathbb{U} & =-\hat{p}_{0,\psi}\mathbb{B}_{-1}+\gamma^{u}\mathbb{B},\label{uF:0}\\
\mathbb{H} & =-\frac{5}{2}\hat{T}_{0,\psi}\mathbb{B}_{-1}+\gamma^{h}\mathbb{B},\label{hF:0}
\end{align}
where $\gamma^{u}$ and $\gamma^{h}$ are expansion coefficients for
the null space of $\mathsf{D}^{0+}$ ($\mathsf{D}^{0+}\mathbb{B}=0)$,
and $\mathbb{B}=\left(B/B_{0}\right)_{\mathtt{F}}$. Combining Eq.~(\ref{HF:})
with (\ref{hF:0}), we have 
\begin{equation}
\mathsf{D}\mathbb{T}=\gamma^{u}\mathbb{F}^{u}+\gamma^{h}\mathbb{F}^{h}+\hat{p}_{0,\psi}\mathbb{F}^{p}+\hat{T}_{0,\psi}\mathbb{F}^{T},\label{DT:}
\end{equation}
where
\begin{align}
\mathbb{F}^{u} & =-\mathsf{K}^{hh,-1}\mathsf{K}^{h\pi}\mathsf{D}_{W}\mathbb{B},\nonumber \\
\mathbb{F}^{h} & =\mathsf{K}^{hh,-1}\mathbb{B},\nonumber \\
\mathbb{F}^{p} & =-\mathsf{K}^{hh,-1}\left(\mathbb{H}^{p_{0}}-\mathsf{K}^{h\pi}\mathsf{D}_{W}\mathbb{B}_{-1}\right),\nonumber \\
\mathbb{F}^{T} & =-\mathsf{K}^{hh,-1}\left(\mathbb{H}^{T_{0}}+\frac{5}{2}\mathbb{B}_{-1}\right),\label{Fco}
\end{align}
Combining Eq.~(\ref{SF:}) with Eq.~(\ref{10F}) and using Eqs.~(\ref{uF:0})
and (\ref{DT:}), we have
\begin{equation}
\mathsf{D}\mathbb{N}+\mathsf{D}\mathbb{T}=\gamma^{u}\mathbb{G}^{u}+\gamma^{h}\mathbb{G}^{h}+\hat{p}_{0,\psi}\mathbb{G}^{p}+\hat{T}_{0,\psi}\mathbb{G}^{T}\label{DP:}
\end{equation}
where
\begin{align}
\mathbb{G}^{u} & =-\mathsf{D}^{1+}\left(\mathsf{K}^{\pi h}\mathbb{F}^{u}+\mathsf{K}^{\pi\pi}\mathsf{D}_{W}\mathbb{B}\right),\nonumber \\
\mathbb{G}^{h} & =-\mathsf{D}^{1+}\mathsf{K}^{\pi h}\mathbb{F}^{h},\nonumber \\
\mathbb{G}^{p} & =-\mathsf{D}^{1+}\left(\mathbb{S}^{p_{0}}+\mathsf{K}^{\pi h}\mathbb{F}^{p}-\mathsf{K}^{\pi\pi}\mathsf{D}_{W}\mathbb{B}_{-1}\right),\nonumber \\
\mathbb{G}^{T} & =-\mathsf{D}^{1+}\left(\mathbb{S}^{T_{0}}+\mathsf{K}^{\pi h}\mathbb{F}^{T}\right).\label{Gco}
\end{align}
The temperature and density can be obtained by inverting the nonsingular
part of $\mathsf{D}$ in Eqs.~(\ref{DT:}) and (\ref{DP:}). The
null space of $\mathsf{D}$ is spanned by $[\varphi_{(0)}]_{\mathtt{F}}$,
which corresponds to the constant term in the Fourier series. Since
the lowest-order density ($n_{0})$ and temperature $(T_{0})$ are
constant, we set $n_{(0)}=0$ and $T_{(0)}=0$ without loss of generality.
From the first row corresponding to the constant $(0)$ Fourier mode,
\begin{align}
0 & =\gamma^{u}\mathbb{F}_{(0)}^{u}+\gamma^{h}\mathbb{F}_{(0)}^{h}+\hat{p}_{0,\psi}\mathbb{F}_{(0)}^{p}+\hat{T}_{0,\psi}\mathbb{F}_{(0)}^{T},\label{aF}\\
0 & =\gamma^{u}\mathbb{G}_{(0)}^{u}+\gamma^{h}\mathbb{G}_{(0)}^{h}+\hat{p}_{0,\psi}\mathbb{G}_{(0)}^{p}+\hat{T}_{0,\psi}\mathbb{G}_{(0)}^{T},\label{aG}
\end{align}
we can determine the integration constants $\gamma^{u}$ and $\gamma^{h}$,
\begin{equation}
\left(\begin{array}{c}
\gamma^{u}\\
\gamma^{h}
\end{array}\right)=-\left(\begin{array}{cc}
\mathbb{F}_{(0)}^{u} & \mathbb{F}_{(0)}^{h}\\
\mathbb{G}_{(0)}^{u} & \mathbb{G}_{(0)}^{h}
\end{array}\right)^{-1}\left(\begin{array}{cc}
\mathbb{F}_{(0)}^{p} & \mathbb{F}_{(0)}^{T}\\
\mathbb{G}_{(0)}^{p} & \mathbb{G}_{(0)}^{T}
\end{array}\right)\left(\begin{array}{c}
\hat{p}_{0,\psi}\\
\hat{T}_{0,\psi}
\end{array}\right).\label{ga:}
\end{equation}
Then Eqs.~(\ref{uF:0}) and (\ref{hF:0}) with the constants obtained
in Eq.~(\ref{ga:}) agree with the corresponding column vectors of
the solution (\ref{M:sol}). Note that the heat flux obtained here
is not a closure and satisfies $\nabla\cdot\mathbf{h}=0$. 

Before concluding this section, a few remarks are in order. First,
Eqs.~(\ref{00F}) and (\ref{01F}) are equivalent to $\nabla\cdot\left(n_{0}\mathbf{V}_{1}\right)=0$
and $\nabla\cdot\mathbf{h}=0$. Inserting the lowest order solutions
$\mathbf{V}_{1\perp}=(1/qB^{2})\mathbf{B}\times\nabla p_{0}$ and
$\mathbf{h}_{\perp}=(5p_{0}/2qB^{2})\mathbf{B}\times\nabla T_{0}$
obtained from $\nabla p_{0}-n_{0}q\mathbf{V}_{1}\times\mathbf{B}/m=0$
and $(5/2)p_{0}\nabla T_{0}-q\mathbf{h}\times\mathbf{B}=0$, one can
derive $\hat{u}=-\hat{p}_{0,\psi}B_{0}/B+\gamma^{u}B/B_{0}$ and $\hat{h}=-5\hat{T}_{0,\psi}B_{0}/2B+\gamma^{h}B/B_{0}$
where $\gamma^{u}$ and $\gamma^{h}$ are integration constants. Second,
$\mathbb{F}^{p}$ and $\mathbb{G}^{p}$ vanish when ion-electron collisions
are ignored. By setting $f_{1}=g+F$, Eq.~(\ref{dke1}) becomes $v_{\|}\partial_{\|}g=C(g)+C(F)$.
Note that the $\hat{p}_{0,\psi}$ term in $C(F)=C(F,f^{0})+C(f^{0},F)$
vanishes due to momentum conservation and does not affect $g$. Therefore
the term $\hat{p}_{0,\psi}$ contributes only to the flow velocity
moment of $f_{1}$ and hence $\mathbb{F}^{p}$ in Eq.~(\ref{DT:})
and $\mathbb{G}^{p}$ in Eq.~(\ref{DP:}) must vanish. Third, in
the closure calculation, the $\hat{p}_{0,\psi}$ drive appears in
$W_{\theta}$ of the viscosity equation and affects closure quantities.
However, the $\hat{p}_{0,\psi}$ term in $V_{1\|}$ of $W_{\theta}$
exactly cancels the $\hat{p}_{0,\psi}$ term in $\mathbf{V}_{1\perp}$
of $W_{\theta}$ making $n_{1}$ and $T_{1}$ independent of the $\hat{p}_{0,\psi}$
drive. Fourth, for an electron-ion plasma $(a,b)=(\mathrm{e},\mathrm{i})$
and $(\mathrm{i},\mathrm{e})$, the $\hat{p}_{a0,\psi}$ and $\hat{p}_{b0,\psi}$
drives do not vanish in the collision operator $C(F_{a},f_{b}^{0})+C(f_{a}^{0},F_{b})$
for the $g_{a}$ equation and do affect $g_{a}$ unless $V_{1a\|}=V_{1b\|}$. 

\section{Conclusion and future work\label{sec:con}}

We have demonstrated how to solve the drift kinetic equation using
the general moment equations to obtain transport and closure relations.
Using the moment-Fourier method developed here, one can directly solve
a full set of parallel moment equations equivalent to the drift kinetic
equation for fluid variables (density, flow velocity, and temperature)
and/or fluxes (particle flux, electric current, heat flux, etc.).
The solution moments can be used to construct the distribution function
that is the solution of the drift kinetic equation. One can also solve
the non-Maxwellian moment equations to express parallel closures in
terms of fluid variables. The closures can be combined with linearized
fluid equations to reproduce the fluid variables and/or fluxes obtained
from the full set of parallel moment equations. More importantly,
the closures can be utilized to advance a system of fluid equations
in numerical simulations with nonlinear terms kept when nonlinear
effects are significant. Note that the drift kinetic equation yields
only linearized fluid equations by nature, e.g. Eqs.~(\ref{me00})-(\ref{me10}),
and hence cannot capture the nonlinear effects. 

While the formalism developed here is only applied in the case of
a single component plasma in a circular axisymmetric magnetic field,
it can be generalized to a multi-component plasma in a tokamak with
arbitrarily shaped nested flux surfaces. As long as the magnetic field
is Fourier-expandable, the moment-Fourier approach developed here
is applicable. For a multi-component plasma, the collisional heating
and friction terms, respectively, will modify Eqs.~(\ref{me01})
and (\ref{me10}). The collision terms introduce couplings of temperatures
and flow velocities between unlike species and, as a result, the $dp_{0}/d\psi$
term will affect all other fluid and closure moments as remarked at
the end of Sec.~\ref{sec:FE=000026C}. Although ion-electron collisions
in the ion theory are ignored based on the small-mass-ratio approximation
in the existing theories (including this work), the momentum and energy
conservations require those terms in the ion fluid equations. These
effects can be investigated by solving coupled moment equations with
the Fourier method. The transport and closure relations for an electron-ion
plasma will be presented in the near future.

The moment-Fourier method developed here is applicable to a plasma
with an arbitrary Knudsen number in a general magnetic field, as long
as convergence can be achieved by increasing the number of moments
and Fourier modes. In the high-collisionality limit, $B/B^{\theta}\lambda_{\mathrm{C}}\ll1$,
the closure coefficients $\mathsf{K}^{\alpha\beta}$ in Eqs.~(\ref{HF:})
and (\ref{SF:}) reproduce the corresponding Braginskii closure coefficient~\citep{Braginskii1965,Ji2015H}.
In the small inverse aspect ratio limit, $\epsilon\ll1$, the $\mathsf{K}^{\alpha\beta}$
reproduce the corresponding integral closure~\citep{Ji2017LH}. In
principle, the moment-Fourier solutions are practically exact once
convergence is achieved. The necessary numbers of moments and Fourier
modes, respectively, increase as the Knudsen number and the inverse
aspect ratio increase. In practice, the moment approach is limited
by the accuracy of the inverse matrix in Eqs.~(\ref{M:sol}) and
(\ref{M':sol}). For low collisionality $\mathtt{n_{F}K_{0}}\gtrsim10^{4}$,
the required matrix dimension for convergence is $LKF\gtrsim10^{6}$,
and the inverse matrix becomes inaccurate due to a large condition
number, even with the exact null space eliminated in the case of Eq.~(\ref{M:sol}).
For low collisionality, the drift kinetic equation may be solved numerically.
However, in the collisionless limit, we find that the drift kinetic
equation should be solved analytically for accurate closure and transport
relations. The results in the collisionless limit will be presented
in the near future, too. It is also notable that the finite element
basis used in Refs.~\citep{Jepson2021e4} and \citep{Spencer2022e4}
makes the convergence faster than the Legendre polynomial basis. 

Since the computational effort to calculate the convergent closures
is tremendous when the effective collisionality is low, it may be
impractical to compute the closures during a fluid simulation. For
practical applications, we plan to develop explicit formulas of closures
which can be expressed in terms of magnetic field parameters, $\epsilon$
for a circular geometry or Fourier components for a general magnetic
field. The explicit expressions of closures can be developed for practical
values of $\epsilon\lesssim0.4$ (at the edge of the ITER tokamak)
and $\mathtt{n_{F}K_{0}}\lesssim10^{4}$ (at the core of ITER). Once
the closures have been obtained for the magnetic field parameters,
they can be conveniently used without time-consuming moment calculations.
Furthermore, calculating $\gamma^{u}$ in Eq.~(\ref{uF:0}) will
be performed for general $\epsilon$ and collisionality of interest
for a quantitative analysis of convergence depending on the number
of moments and Fourier modes. 

\section*{Data availability statement}

The data that support the findings of this study are available upon
request from the authors.
\begin{acknowledgments}
The research was supported by the U.S. DOE under Grant Nos. DE-SC0022048
and DE-FG02-04ER54746 and by National R\&D Program through the National
Research Foundation of Korea (NRF) funded by Ministry of Science and
ICT (2021M3F7A1084419).
\end{acknowledgments}

\end{document}